\documentclass{article}
\usepackage{amsfonts}
\usepackage{amsmath}
\usepackage{amssymb}
\begin{document}
\title{Entropic criterion for model selection\small\thanks{Accepted for publication 
in Physica A, 2006}}

\author{Chih-Yuan Tseng\small\thanks{%
E-mail: richard@pooh.phy.ncu.edu.tw}\\
%EndAName
Computational Biology Laboratory\\
Department of Physics \\
National Central University, Chungli, Taiwan 320}
\date{}
\maketitle

\begin{abstract}
Model or variable selection is usually achieved
through ranking models according to the increasing order of preference.
One of methods is applying Kullback-Leibler distance or relative entropy 
as a selection criterion. Yet that will raise two
questions, why uses this criterion and are there any other criteria. Besides,
conventional approaches require a reference prior, which is usually
difficult to get. Following the logic of inductive inference proposed 
by Caticha \cite{Caticha04}, we show relative entropy to be a unique
criterion, which requires no prior information and can be applied
to different fields. We examine this criterion by considering
a physical problem, simple fluids, and results are promising.
\end{abstract}



\section{Introduction}

Model or variable selection in process of data analysis is usually achieved
by ranking models according to the increasing order of preference. Several
methods rooted in this concept such as P-values, Bayesian, and
Kullback-Leibler distance method, etc., are some popular examples to provide
pertinent selection criteria. P-values method selects model by comparing
probability of model given a null model and experimental data sets to a
threshold value assessed from same data sets \cite{Raftery95}. Yet since
this method is restricted to two models and required ad hoc rules to assess
threshold value, people has developed Bayesian approaches to overcome these
defects (\cite{Raftery95}, \cite{Weiss95}, \cite{Raftery97}, \cite{Kieseppa00}, \cite{Forbes03}). 
The Bayesian method applies Bayes theorem to
update our beliefs and uncertainty about models from prior distributions
generated from some prior modeling rules first. A preferable model,
thereafter, is chosen according to Bayes factor, ratio of posterior
distributions of different models. Bayesian Information Criterion (BIC) is
one of most popular model selection criteria (\cite{Raftery95}, 
\cite{Kieseppa00}, \cite{Forbes03}). Yet all of
these methods require prior information generated from some ad hoc prior
modeling rules that suits people's need.

Aside from Bayesian framework, people also has developed relative entropy,
mutual information, or Kullback-Leibler distance based approach (\cite
{Bonnlander94}, \cite{Dupuis03}). Kullback-Leibler distance measures differences between model
and a reference prior for interested system. The decreasing Kullback-Leibler
distance then suggests the increasing order of preference of models. A
selection method proposed by Dupuis and Robert on variable selection \cite
{Dupuis03} is based on the evaluation of Kullback-Leibler distance between
full model described by complete set of variables for interested system
 and it's approximations, submodels, described by subset of variables.
Given prior information on full model, preferred
submodel is selected when it's
Kullback-Leibler distance reaches a threshold value, estimated by
experiences. Since submodels are projections of full model, there is no need
the prior modeling rule to generate prior distribution for submodel. Yet
one still requires prior information on full model. Moreover, it remains
questionable to apply Kullback-Leibler distance as a selection criterion even
though Dupuis and Robert argued that it is a common choice for information
theoretic and intrinsic considerations and computational reasons. Besides,
the choice is made because of it's properties of transitivity and additivity
that relate to theory of generalized linear models \cite{Dupuis03} attempted to apply
in breast cancer studies.
Our first goal of this work is to answer questions of why is
Kullback-Leibler distance not any other criteria. Are there any other
entropy based criteria for model selection? Afterward, we shall develop an
entropy based method to provide ranking scheme for model selection that is
free from difficulties encountered in conventional entropic studies. The
strategy is closely following the logic of inductive inference proposed by \cite{Caticha04}
that is to generalize method of maximum entropy (ME) from Jaynes's version
of probability distribution assignments \cite{Jaynes57} to be a tool of
inductive inference initiated by \cite{ShoreJohnson80}
and \cite{Skilling88}. 
This logic differs in one remarkable way from the manner that has in the past been followed in setting up physical theories for example. Normally one starts by establishing a
mathematical formalism, and then one tries to append an interpretation to
it. This is a very difficult problem; it has affected the development of
statistics and statistical physics - what is the meaning of probability
distribution and of entropy. The issue of whether the proposed
interpretation is unique, or even whether it is allowed, always remains a
legitimate objection and a point of controversy. Our procedure is in the
opposite order, we first decide what are we talking about and what is our
goal, namely, selection criterion for ranking scheme, and only afterward we design the appropriate
mathematical formalism, the issue of what is the meaning of probability
distributions and of entropy will then never arise.

Based on Caticha's logic of inductive inference, we shall derive and present
the entropic criterion in next section first. It will show relative entropy
to be a unique criterion for model selection. We thereafter examine this
criterion by considering a complicated physical problem, simple fluids. It
has become a well matured field after almost three decades. People has
developed many approximation models to study and interpret fluid's
properties. Since we have rich theoretical and experimental knowledge of
simple fluids from conventional studies to rank those approximation models, it shall provide us a
conceivable benchmark for our investigations. Three approximation models,
mean field, hard-sphere and improved hard-sphere approximation are
considered and briefly presented in section 3-1. We then apply entropic
criterion to rank these three models. Detail calculations of entropic
criterion and comparison against results inferred from conventional analysis
are shown in section 3-2. A summary of our discussions is listed in section 4.

\section{Entropic criterion for model selection}

As mentioned, the selection of one model from within a group of models is
achieved by ranking those models according to increasing \emph{preference}.
Before we address the issue of what it is that makes one model preferable
over another we note that there is one feature we must impose on any ranking
scheme. The feature is transitivity: if model 1 described by distribution $%
p_{1}$ is preferred over model 2 described by distribution $p_{2}$, and $%
p_{2}$ is preferred over $p_{3}$, then $p_{1}$ is preferred over $p_{3}$.
Such transitive rankings are implemented by assigning to each $p(x)$ a real
number $S[p]$ which we call the ``entropy''
of $p$. The numbers $S[p]$ are such that if $p_{1}$ is preferred over $p_{2}$%
, then $S[p_{1}]>S[p_{2}]$.

Next we determine the functional form of $S[p]$. The basic strategy 
\cite{Skilling88} is that (1) if a general rule exists, then it must apply to
special cases; (2) if in a certain special case we know which is the best
model, then this knowledge can be used to constrain the form of $S[p]$; and
finally, (3) if enough special cases are known, then $S[p]$ will be
completely determined.
The known special cases are called the 
``axioms'' of ME and they reflect the conviction that one
should not change one's mind frivolously, that whatever information was
codified in probability distribution $p(x)$ is important. Three axioms and
their consequences are listed below. Detailed proofs are given in \cite%
{Caticha04}.

\textbf{Axiom 1: Locality}. \emph{Local information has local effects.} If
the constraints that define the probability distribution do not refer to a
certain domain $D$ of the variable $x$, then the conditional probabilities $%
p(x|D)$ need not be revised. The consequence of the axiom is that
non-overlapping domains of $x$ contribute additively to the entropy: $%
S[p]=\int dx\,F(p(x))$ where $F$ is some unknown function.

\textbf{Axiom 2: Coordinate invariance.} \emph{The ranking should not depend
on the system of coordinates. }The coordinates that label the points $x$ are
arbitrary; they carry no information. The consequence of this axiom is that $%
S[p]=\int dx\,p(x)f(p(x)/m(x))$ involves coordinate invariants such as $%
dx\,p(x)$ and $p(x)/m(x)$, where the function $m(x)$ is a density, and both
functions $m$ and $f$ are, at this point, unknown.
We make a second use of the locality axiom to determine $m(x)$. When there
are no constraints at all and group of different models includes the exact $%
P(x)$ for real system, the selected probability model $p(x)$ should coincide
with $P(x)$; that is, the best probability model $p(x)$ to real system
described by $P(x)$ is $P(x)$ itself. On the contrary it suggests that the
best probability model $p(x)$ to $P(x)$ should be farthest from uniform
distribution $m$. Since exact distribution $P(x)$ is sometimes too
complicated to be useful in practical calculations while uniform
distribution $m$ is free from this difficulty. Thus we shall choose uniform
distribution $m$ to be $m(x)$. At last, we will consider third axiom to
determine function $f$.
\textbf{Axiom 3:\ Consistency for independent subsystems}. \emph{When a
system is composed of subsystems that are believed to be independent it
should not matter whether we treat them separately or jointly.} If we
originally believe that two systems are independent and the constraints
defining the probability distributions are silent on the matter of
correlations, then there is no reason to change one's mind. Specifically, if 
$x=(x_{1},x_{2})$, and the exact distributions for the subsystems, $%
p_{1}(x_{1})$ and $p_{2}(x_{2})$, then the exact distribution for the whole
system should be $p_{1}(x_{1})p_{2}(x_{2})$. This axiom restricts the
function $f$ to be a logarithm.

The overall consequence of three axioms is that the probability distribution 
$p(x)$ should be ranked relative to $m$ according to their (relative)
entropy, 
\begin{equation}
S[p,m]=-\int dx\,p(x)\log \frac{p(x)}{m}=\ln m-\int dx\,p(x)\log p(x)\leq 0.
\label{S[p]}
\end{equation}%
The derivation has singled out $S[p,m]$ as \emph{the unique entropy to be
used for the purpose of ranking probability distributions}. Other
expressions, may be useful for other purposes, but they are not a
generalization from the simple cases described in the axioms above. Notice
that since $p(x)$ is ranked relative to a uniform distribution $m$, which is
independent of models. Thus decreasing $S[p]=-\int dx\,p(x)\log p(x)$ in $%
S[p,m]$ indicates there existing more differences between probability model $%
p(x)$ and uniform distribution $m$. Namely, $p(x)$ that is farther away from
uniform distribution carries more relevant information about real system,
and is more preferable. 
Before applying entropic criterion to real problems, a summary of our
derivation is given. Based on logic of inductive inference, the answer to
questions raised earlier becomes obviously. The use of relative entropy for
selection criterion is just what we design to achieve, and needs no further
interpretation. Besides, since this criterion is designed based on 
probability models, it will accommodate to all kinds of probabilistic problems. 

\section{A physical problem: simple fluids}
\subsection{Approximation models for simple fluids}
We shall examine proposed entropic criterion by considering a complicated
problem in physics, simple fluids (reviews of simple fluids can be found in
\cite{Barker76}, \cite{Hansen86}, and \cite{Kalikmanov02}), in this section. Suppose a
simple fluid with density $\rho $ and volume $V$ is composed of $N$ single
atom molecules. This fluid is described by the Hamiltonian 
\begin{equation}
H(q_{N})=\sum_{i=1}^{N}\,\frac{p_{i}^{2}}{2m}+U\quad \text{with}\quad
U=\sum_{i>j}^{N}u(r_{ij})\,,  \label{Actual-H}
\end{equation}%
where $q_{N}=\{p_{i},r_{i};\;i=1,...,N\}$ and the many-body interactions are
approximated by a Lennard-Jones pair interaction, $u(r_{ij})=4\epsilon
\left( \left( \sigma /r_{ij}\right) ^{12}-\left( \sigma /r_{ij}\right)
^{6}\right) $ where $r_{ij}=\left\vert r_{i}-r_{j}\right\vert $ and $%
\epsilon $ and $\sigma $are Lennard-Jones parameters. The probability that
the positions and momenta of the molecules lie within the phase space volume 
$dq_{N}=\frac{1}{N!\,h^{3N}}\prod_{i=1}^{N}d^{3}p_{i}d^{3}r_{i}$, 
$P(q_{N})\,dq_{N}$ is given by%
\begin{equation}
P(q_{N})\,dq_{N}=\frac{1}{Z}\exp -\beta H(q_{N})\text{ and }Z=\int
dq_{N}\exp -\beta H(q_{N})\text{ },  \label{P(real)}
\end{equation}%
where $\beta =\frac{1}{k_{B}T}$ and $k_{B}$ is Boltzmann constant.
Since there are $N\left( N-1\right) /2$ pair interactions, integration of
partition function $Z$ in Eq.(\ref{P(real)}) is impossible to accomplish. 
$P(q_{N})$ is useless in practical calculations. One strategy to bypass this
problem is constructing approximation models that are described by tractable
probability distributions. Several approximation models are, therefore,
developed according to researchers's knowledge and experiences in the
studies of simple fluids in last three decades. Yet we shall consider only
three approximation models, mean field from \cite{Tseng}, hard-sphere from \cite{mansorri69} or
\cite{Tseng2} and
improved hard-sphere approximation from \cite{Tseng2} that are briefly presented
in the following to demonstrate the use of entropic criterion for model
selection.

\textbf{Mean field approximation:} \emph{Mean field approximation
drastically approximates complicated long range interactions }$u(r_{ij})$%
\emph{\ by an optimal mean field }$v_{mf}(r_{ij})$, \emph{which is
determined by ME\ method from} \cite{Tseng}. Probability distribution given by
mean field approximation%
\begin{equation}
P_{mf}(q_{N};\beta ,\lambda )=\frac{1}{Z_{mf}}\exp -\beta \left[
H_{mf}(q_{N})+\int d^{3}r\,\,\lambda (r)\hat{n}(r)\right] \,,
\label{trial P a}
\end{equation}%
where%
\begin{equation}
H_{mf}(q_{N})=\sum_{i=1}^{N}\,\frac{p_{i}^{2}}{2m}+\int d^{3}r\,\,v_{mf}(r)%
\hat{n}(r)\,,
\end{equation}%
and $\lambda (r)$ are Lagrange multipliers that enforce the constraint on
the expected density $\langle \hat{n}(r)\rangle $ at each point in space and
the density is 
\begin{equation}
\hat{n}(r)=\sum_{i=1}^{N}\,\delta (r-r_{i})\,.
\end{equation}%
We remark that a constraint on $\langle \hat{n}(r)\rangle $ also constrains
the expected number of particles 
\begin{equation}
N=\langle N\rangle =\int d^{3}r\,\langle \hat{n}(r)\rangle \,.
\end{equation}

It is convenient to absorb the mean field $v_{mf}(r)$ and the multiplier
field $\lambda (r)$ into a single potential $V(r)=v_{mf}(r)+\lambda (r)$,
the partition function $Z_{mf}$ is 
\begin{equation}
Z_{mf}=\sum_{N=0}^{\infty }\int dq_{N}\,\exp -\beta \left[ \sum_{i=1}^{N}\,%
\frac{p_{i}^{2}}{2m}+V(r)\right] \,\overset{\text{def}}{=}e^{-\beta
\Omega _{mf}(\beta ,\lambda )}\,,  \label{partition-mf}
\end{equation}%
so that 
\begin{equation}
\Omega _{mf}[T,\lambda ]=-\frac{1}{\beta \Lambda ^{3}}\int
d^{3}r\,\,e^{-\beta V(r)}=-\frac{1}{\beta }\int d^{3}r\,\,n_{mf}\left(
r\right) =-\frac{N}{\beta }\,,  \label{Omega0}
\end{equation}%
where$\quad \Lambda =\left( \frac{\beta h^{2}}{2\pi m}\right) ^{1/2}$ and
the expected density is 
\begin{equation}
\langle \hat{n}(r)\rangle _{mf}=\frac{\delta \Omega _{MF}}{\delta \lambda (r)%
}=\frac{\delta \Omega _{MF}}{\delta V(r)}\overset{\text{def}}{=}n_{mf}(r)
\label{dOmega/dV}
\end{equation}%
or 
\begin{equation}
n_{mf}(r)=\frac{e^{-\beta V(r)}}{\Lambda ^{3}}\,.  \label{n0}
\end{equation}%
Furthermore, according to Percus-Yevick approximation, one can introduce a
useful quantity, radial distribution function $g_{mf}(r)=n_{mf}(r)/\rho $
that measures probability of observing particle at distance $r$ while
another particle at origin. It gives information of liquid structures that
can be measured directly through x-ray and neutron diffraction experiments.
Notice that when two particles are not correlated, radial distribution
function $g_{mf}(r)=1$. $g_{mf}(r)$ is vanished when two atoms are repelled.

\textbf{Hard-sphere approximation:} \emph{One replaces short-range repulsion
by hard-sphere potential with an optimal hard-sphere diameter }$\bar{r}_{d}$,%
\emph{\ which is determined by ME\ method as well } \cite{Tseng2}.
Probability distribution given by hard-sphere approximation is 
\begin{equation}
P_{hs}(q_{N}\left\vert \bar{r}_{d}\right. )=\frac{1}{Z_{hs}}e^{-\beta
H_{hs}(q_{N}\left\vert \bar{r}_{d}\right. )}\,,  \label{Phs}
\end{equation}%
where the Hamiltonian is 
\begin{equation}
H_{hs}(q_{N}\left\vert \bar{r}_{d}\right. )=\sum_{i=1}^{N}\,\frac{p_{i}^{2}}{%
2m}+U_{hs}  \label{Hhs}
\end{equation}%
with 
\begin{equation}
U_{hs}=\sum_{i>j}^{N}u_{hs}(r_{ij}|\bar{r}_{d})~,\text{ }
\end{equation}%
where 
\begin{equation}
u_{hs}(r\left\vert \bar{r}_{d}\right. )=\left\{ 
\begin{array}{ccc}
0 & \text{for} & r\geq \bar{r}_{d} \\ 
\infty & \text{for} & r<\bar{r}_{d}%
\end{array}%
\right. \text{ }.
\end{equation}%
The partition function and the free energy $F_{hs}(T,V,N\left\vert \bar{r}%
_{d}\right) $ obtained by Percus-Yevick approximation \cite{Hansen86} are 
\begin{equation}
Z_{hs}=\int dq_{N}\,\text{\ }e^{-\beta H_{hs}(q_{N})}\,\overset{\text{def}}{=%
}e^{-\beta F_{hs}(\beta ,V,N\left\vert \bar{r}_{d}\right. )}\,
\label{partition-hs}
\end{equation}%
and
\begin{equation}
F_{hs}(\beta ,V,N\left\vert \bar{r}_{d}\right. )=Nk_{B}T\left[ -1+\ln \rho
\Lambda ^{3}+\frac{4\bar{\eta}-3\bar{\eta}^{2}}{\left( 1-\bar{\eta}\right)
^{2}}\right] ,\text{ }  \label{HS-Free energy}
\end{equation}%
where the packing fraction, $\bar{\eta}\overset{\text{def}}{=}\frac{1}{6}\pi
\rho \bar{r}_{d}^{3}\quad $with$\quad \rho =\frac{N}{V}~.$

\textbf{Improved hard-sphere approximation:} Although there are several
improved hard-sphere models like Barker and Henderson \cite{Barker76} and WCA 
theories \cite{WCA71} 
etc., they were not constructed by
probability models directly, and are inappropriate in this investigation. We
consider another improved approximation model obtained from method of ME
that is a probability model and has been proved to be competitive to those
theories \cite{Tseng2}. The crux of this model is that we consider whether
the correct choice should have been some other value $r_{d}=\bar{r}%
_{d}+\delta r$ rather than the optimal $r_{d}=\bar{r}_{d}$ in original
hard-sphere approximation. As discussed in \cite{Tseng2} this is a question
about the probability of $r_{d}$, $P_{d}(r_{d})$. Thus, we are uncertain not
just about $q_{N}$ but also about $r_{d}$ and what we actually seek is the
joint probability of $q_{N}$ and $r_{d}$, $P_{J}(q_{N},r_{d})$. Once this
joint distribution is obtained our best assessment of the distribution of $%
q_{N}$ is given by the marginal over $r_{d}$,%
\begin{eqnarray}
\bar{P}_{hs}(q_{N}) &\overset{\text{def}}{=}&\int dr_{d}\text{ }%
P_{J}(q_{N},r_{d})  \notag \\
&=&\int dr_{d}\text{~}P_{d}(r_{d})P_{hs}(q_{N}|r_{d}).\text{ }
\label{Pbar(q)}
\end{eqnarray}%
where the distribution of diameters is given by 
\begin{equation}
P_{d}(r_{d})dr_{d}=\frac{e^{\mathcal{S}\left[ P_{hs}|P\right] }}{\zeta }%
\gamma ^{1/2}\left( r_{d}\right) dr_{d}=\frac{e^{-\beta F_{U}}}{\zeta _{U}}%
\gamma ^{1/2}\left( r_{d}\right) dr_{d}\text{ },  \label{Pd(rd)}
\end{equation}%
$\gamma \left( r_{d}\right) =N\pi \rho r_{d}\frac{4+9\eta -4\eta ^{2}}{%
\left( 1-\eta \right) ^{4}}$, $\eta \overset{\text{def}}{=}\frac{1}{6}\pi
\rho r_{d}^{3}$ and the partition functions $\zeta $ and $\zeta _{U}$ are
given by
\begin{equation}
\zeta =e^{\beta F}\zeta _{U}\quad \text{with}\quad \zeta _{U}=\int dr_{d}%
\text{ }\gamma ^{1/2}\left( r_{d}\right) e^{-\beta F_{U}}\text{ ,}
\end{equation}%
and
\begin{equation}
\mathcal{S}\left[ P_{hs}|P\right] =-\int dq_{N}\,P_{hs}(q_{N}|r_{d})\log 
\frac{P_{hs}(q_{N}|r_{d})}{P(q_{N})}=\beta \left( F-F_{U}\right) \text{ }
\end{equation}%
with $F_{U}=F_{hs}(\beta ,V,N\left\vert r_{d}\right. )+\frac{1}{2}N\rho \int
d^{3}r\,u(r)g_{hs}(r\left\vert r_{d}\right. )$. In addition, one have to
consider proper local fluctuation effect in model to generate correct liquid
structure by requesting $N$ to be effective particle number $N_{eff}$ \
(please refer to \cite{Tseng2} for more discussions). By recognizing that
diameters other than $\bar{r}_{d}$ are not ruled out and that a more honest
representation is an average over all hard-sphere diameters we are
effectively replacing the hard-spheres by a soft-core potential.

\subsection{Discussions}
\subsubsection{Entropic criterion analysis}
Now, we first implement proposed entropic criterion to rank three
approximation models for simple fluids in last section before presenting the
actual ranking scheme inferred from detail analysis of liquid structures and
thermodynamical properties obtained by these approximations against to
computer simulations and experimental data. According to proposed entropic
criterion, ranking scheme is obtained by calculating entropy $S[p_{i}]$\ of
probability distributions $p_{i}$ of i$^{\text{th}}$ approximations.
Substituting Eq.(\ref{trial P a}) into $S[p_{i}]$, entropy per particle
number of mean field approximation $P_{mf}$ 
\begin{eqnarray}
S[P_{mf}]/Nk_{B} &=&-\beta \Omega _{mf}(\beta ,\lambda )/N+\frac{3}{2}+\frac{%
\beta }{N}\int d^{3}r\,\,V(r)n_{mf}(r)  \notag \\
&=&\frac{5}{2}-\log \rho \Lambda ^{3}-\frac{1}{V}\int d^{3}r\,g_{mf}(r)\log
g_{mf}(r)  \label{S[pmf]}
\end{eqnarray}%
is obtained with the help of Eq.(\ref{partition-mf}), (\ref{Omega0}), (\ref%
{dOmega/dV}), and $g_{mf}(r)=n_{mf}(r)/\rho $. Next, entropy of hard-sphere
approximation $P_{hs}$ is calculated by differentiating free energy $F_{hs}$%
, Eq.(\ref{HS-Free energy}), with respect to temperature,%
\begin{eqnarray}
S[P_{hs}]/Nk_{B} &=&-\beta F_{hs}(\beta ,V,N\left\vert \bar{r}_{d}\right.
)/N+\frac{3}{2}  \notag \\
&=&\frac{5}{2}-\log \rho \Lambda ^{3}-\left[ \frac{4\bar{\eta}-3\bar{\eta}%
^{2}}{\left( 1-\bar{\eta}\right) ^{2}}\right] \text{ .}  \label{S[phs]}
\end{eqnarray}%
At last, entropy of probability distribution $\bar{P}_{hs}$ given by
improved hard-sphere approximation is obtained by substituting Eq.(\ref%
{Pd(rd)}) and (\ref{Phs}) into Eq.(\ref{Pbar(q)}) first. Because $\exp
-\beta U_{hs}$ is a Heaviside step function, one can write spatial part of $%
\bar{P}_{hs}$ as%
\begin{equation}
\bar{P}_{hs}^{\prime }(r)=\int_{0}^{r}dr_{d}\text{ }P_{d}(r_{d})\exp \beta
F_{\eta }(r_{d})\text{ with }\beta F_{\eta }(r_{d})=N\left[ \frac{4\eta
-3\eta ^{2}}{\left( 1-\eta \right) ^{2}}\right] \text{ }
\label{Pbar(q)prime}
\end{equation}%
Since $P_{d}(r_{d})$ is vanished when $r>r_{t}$ and $r<r_{b}$, integrating $%
P_{d}(r_{d})\exp \beta F_{\eta }(r_{d})$ from zero to $r>r_{t}$ in Eq.(\ref%
{Pbar(q)prime}) will give a constant value, which defines a new quantity $%
\beta \bar{F}_{\eta }$. Therefore, $S[\bar{P}_{hs}]$ is given by%
\begin{eqnarray}
S[\bar{P}_{hs}]/N_{eff}k_{B} &=&\frac{5}{2}-\log \rho \Lambda ^{3}-\beta 
\bar{F}_{\eta }  \notag \\
&&-\frac{1}{N_{eff}V^{\prime }}\int_{r_{b}}^{r_{t}}d^{3}r\,\,\bar{P}%
_{hs}^{\prime }(r)\log \bar{P}_{hs}^{\prime }(r),  \label{S[iPhs]}
\end{eqnarray}%
where $V^{\prime }=$ $\int_{0}^{\infty }d^{3}r\,\,\bar{P}_{hs}^{\prime }(r)$.
Next, we compare values of Eq.(\ref{S[pmf]}), (\ref{S[phs]}), and (\ref%
{S[iPhs]}). The only difference between entropies of mean field Eq.(\ref{S[pmf]}%
) and of hard-sphere approximation Eq.(\ref{S[phs]}) is the third term
contributed by potential part. Since radial distribution function $g_{mf}(r)$
in Eq.(\ref{S[pmf]}) is vanished within range of strong repulsive forces, $%
r=0$ and $r_{l}$, and becomes one after $r\geq r_{u}$, this result leads to
a constant integration of third term that is far smaller than total fluid
volume $V$. Therefore, entropy of $P_{mf}$ is approximated to%
\begin{equation}
S[P_{mf}]/Nk_{B}\approx \frac{5}{2}-\log \rho \Lambda ^{3}\text{ }.
\end{equation}%
One thereafter has the inequality equation,
\begin{equation}
S[P_{hs}]/Nk_{B}<S[P_{mf}]/Nk_{B}\text{ ,}
\end{equation}%
hard-sphere approximation is preferred over mean field approximation.
Now consider entropy of approximations $P_{hs}$ and $\bar{P}%
_{hs}(q_{N_{eff}})$. Numerical calculations of third term in Eq.(\ref{S[phs]}%
) and sum of last two terms in Eq.(\ref{S[iPhs]}) with three different fluid's
densities and temperatures are shown in Table.\ref{tab1} denoted by HS3 and IHS3
respectively as examples. These numerical values shows IHS3 to be smaller than HS3, 
namely, $S[\bar{P}_{hs}]/N_{eff}k_{B}<S[P_{hs}]/Nk_{B}$, and suggest an expected result that improved hard-sphere approximation is preferred over hard-sphere approximation.
Therefore, the complete ranking scheme of these three approximations is
\begin{equation}
S[\bar{P}_{hs}]/N_{eff}k_{B}\leq S[P_{hs}]/Nk_{B}<S[P_{mf}]/Nk_{B}<0\text{ },
\label{ranking}
\end{equation}%
where the equality in $S[\bar{P}_{hs}]\leq S[P_{hs}]$ will hold when $%
N_{eff} $ increase to equal to total particle number, which results in
improved hard-sphere approximation to reduce to hard-sphere approximation as
discussed in \cite{Tseng2}.

\begin{table}
 \caption{\label{tab1} This table lists values of third term in Eq.(\ref{S[phs]}%
) denoted by HS3 and sum of last two terms in Eq.(\ref{S[iPhs]}) denoted by IHS3 for 
three different fluid densities $\rho$ and temperatures T.}
\centering
\fbox{%
\begin{tabular}{|c|c|c|c|c|c|c|}
\hline
$\rho \sigma ^{3}$ & \multicolumn{2}{|c}{0.55} & \multicolumn{2}{|c}{0.65} & 
\multicolumn{2}{|c|}{0.7} \\ \hline
& HS3 & IHS3 & HS3 & IHS3 & HS3 & IHS3 \\ \hline
T=107.82K & -1.758 & -2.284 & -2.297 & -3.182 & -2.603 & -3.682 \\ \hline
161.73 & -1.614 & -2.153 & -2.081 & -2.881 & -2.342 & -3.282 \\ \hline
328.25 & -1.376 & -1.832 & -1.744 & -2.338 & -1.944 & -2.614 \\ \hline
\end{tabular}}
\end{table}

\subsubsection{Conventional analysis}
Alternatively, one can determine the ranking scheme of these three
approximations through exhausting analysis of comparing liquid structures
and thermodynamical properties obtained by these approximations against to
computer simulations and experimental data (please refer to \cite{Tseng} and 
\cite{Tseng2} for detail). Results showed that mean field approximation only
suits for dilute gases and fails to take short-range interaction into
account properly. Contrarily, hard-sphere approximation fails to take
softness of the repulsive core, which results in less satisfactory
prediction of thermodynamic properties at high temperature. Yet hard-sphere
approximation still provides better description of short-range interactions
than mean field approximation does. Furthermore, since improved hard-sphere
approximation attempts to take softness of repulsive core into account
pertinently, results showed such an improvement to be competitive with the
best perturbative theories so far. One, therefore, can rank these three
approximations for simple fluids studies from these analysis as follows,
improved hard-sphere approximation is preferred over hard-sphere
approximation and hard-sphere approximation is preferred over mean field
approximation. This is exactly the same ranking scheme as indicated by Eq.(%
\ref{ranking}) yet it requires more exhausting efforts.

\section{Discussion}
There has been abundant theories proposed to construct robust and efficient model or variable selection criteria. We briefly reviewed the rationale and some shortcomings of these
methods. P-values method \cite{Raftery95} is restricted to two models and
requires some ad hoc rules determining threshold value. Although several Bayesian
methods (\cite{Weiss95}, \cite{Raftery95}, \cite{Raftery97}, \cite{Kieseppa00},
\cite{Forbes03}) are proposed to overcome this shortcoming, it
still requires prior modeling rule to generate prior distribution. Aside
from Bayesian framework, there are relative entropy, mutual information %
or Kullback-Liebler distance methods for the same goal (\cite{Bonnlander94}, \cite{Dupuis03}).
In \cite{Dupuis03}, Dupuis and Robert applied
Kullback-Liebler distance to select submodels, projections of a full model
given a full model for interested system. Yet this approach still requires
prior information on full model and a threshold value. Moreover, it remains
questionable to choose Kullback-Liebler distance as the selection criterion
even though Dupuis and Robert gave some arguments to defend such a choice.
Our first goal is to answer questions of why is Kullback-Liebler distance for
selection criterion and are there any other criteria. Afterward, we propose
entropic criterion to determine ranking scheme given a group of several
models for a system. Following logic of inductive inference proposed by \cite{Caticha04} as
mentioned in introduction, we answer these two questions by
showing relative entropy to be a unique criterion to rank different models
for a system. It is just what we design to achieve, and needs no further
interpretation. Besides, there is no restriction on types of probability
models in this criterion, and it has wide applicability in
all kinds of probabilistic problems. 
Since probability distribution of real system, however, is always
intractable that is useless in practical calculations, we propose to rank
probability models relative to a uniform distribution $m$ instead real
probability distribution to bypass this defect. Thus decreasing relative
entropy $S[p,m]$ indicates increasing preference of models. Notes that it
has no restrictions on numbers of models and requires no ad hoc prior
modeling rules.
At last, we examine this tool by considering a complicated physical problem,
simple fluids in this work. Because people has developed many approximation
models to study simple fluids, and accumulated rich knowledge in the past,
it provides a conceivable benchmark for our investigation. We consider three
approximations models, mean field from \cite{Tseng}, hard-sphere, and improved hard-sphere
approximation from \cite{Tseng2} 
for demonstration. Calculations of entropic criterion of these 
three approximations straightforwardly gives the same ranking scheme, improved hard-sphere
approximation is preferred over hard-sphere approximation and hard-sphere
approximation is preferred over mean field approximation, as
inferred by thoroughly but exhausting analysis based on our own knowledge and
results against to computer simulations and experimental data.

\section*{Acknowledgement}
This work is partially supported by grant NSC-94-2811-M-008-018 from
National Science Council, ROC. Author is grateful many discussions with
colleagues Bao-Zhu Yang and Chien-Chih Chen regarding to practical
applications of model selection method in genome and geophysics problems
respectively.

\end{document}